# Unified force law for granular impact cratering

H. Katsuragi[*] and D. J. Durian

*Department of Physics and Astronomy, University of Pennsylvania, Philadelphia, Pennsylvania 19104-6396, USA*

Experiments on the low-speed impact of solid objects into granular media have been used both to mimic geophysical events [1-5] and to probe the unusual nature of the granular state of matter [6-9]. While the findings are all strikingly different from impact into ordinary solids and liquids, no consensus has emerged regarding the interaction between medium and projectile. Observation that the final penetration depth is a power of the total drop distance was interpreted by a stopping force that is a product of powers of depth and speed [6]. Observation that the penetration depth is linear in initial impact speed was interpreted by a force that is linear in speed [7]. Observation that the stopping time is constant was interpreted by a force that is constant but proportional to the initial impact speed [8]. Observation that depth vs time is a sinusoid for a zero-speed impact was interpreted by a force that is proportional to depth [9]. These four experimental results, as well as their interpretations, would all seem to be in conflict. This situation is reminiscent of high-speed ballistics impact in the 19[th] and 20[th] centuries, when a plethora of empirical rules were proposed [10,11]. To make progress, we developed a means to measure projectile dynamics with 100 nm and 20 μs precision. For a 1-inch diameter steel

---

[*] Permanent address: Department of Applied Science for Electronics and Materials, Kyushu University, Kasuga, Fukuoka 816-8580, Japan

**sphere dropped from a wide range of heights into non-cohesive glass beads, we reproduce prior observations [6-9] either as reasonable approximations or as limiting behaviors. Furthermore, we combine all our data to demonstrate that the interaction between projectile and medium can be decomposed into the sum of a velocity-dependent inertial drag force plus a depth-dependent frictional force. Thus we achieve a unified description of low-speed impact phenomena and show that the response of granular materials, while fundamentally different from that of liquids and solids, can be surprisingly simple.**

To measure dynamics, we use a line-scan digital CCD camera to image a finely-striped transparent rod attached vertically to the top of the projectile (see Methods). The instantaneous speed is the key quantity, deduced from the displacement of the striped pattern between successive frames. The temporal precision is $20\,\mu\text{s}$, set by the $50\,\text{kHz}$ frame rate of the line-scan camera. The position resolution is $100\,\text{nm}$, set by the $3.8\,\mu\text{m/pixel}$ magnification divided by the square-root of the number of pixels. These combine to give a velocity resolution of $0.5\,\text{cm/s}$. Besides measurement fidelity, another advantage of our method is that it applies even to very deep impacts – as long as the rod does not submerge.

Our complete dynamics data set is displayed in Fig.1: position $z$, velocity $v$, and acceleration $a$, vs time $t$, for initial impact speeds $v_o$ ranging from 0 to $-400\,\text{cm/s}$. Time is measured from initial impact; position is measured upwards from the granular surface, opposite to gravity. A striking feature is that, while the final position is approached smoothly, the velocity vanishes abruptly with a discontinuity in acceleration [12]. This is counter to viscous approach to a stable equilibrium, where acceleration vanishes continuously, but it permits the stopping time $t_{stop}$ to be easily gauged from the velocity vs time data. Note that $t_{stop}$ actually decreases with increasing impact speed; surprisingly, deeper penetration requires less time. Evidently, granular matter is very different from ordinary solids and liquids in its resistance to penetration.

We begin analysis by comparing our impact data with the seemingly contradictory trends reported in Refs.[6-9]. First, the stop time is plotted vs $v_o$ in the inset of Fig.1b. As in Ref.[8], $t_{stop}$ appears constant for fast impacts; however, this is only the limiting behaviour since $t_{stop}$ increases for slower impacts. Second, the instantaneous acceleration is plotted vs depth in the inset of Fig.1c for the case $v_o = 0$. As in Ref.[9], $a$ appears equal to $-g$ plus a Coulomb friction term proportional to depth; however, this is only the limiting behaviour since we find in the main plot that the acceleration at $z = 0$ increases with impact speed. Third, the absolute final penetration depth $d$ is plotted in Fig.2 vs both the total drop distance $H = h + d$ and vs the impact speed $v_o = -(2gh)^{1/2}$, where $h$ is the free-fall distance and $g = 980 \, \text{cm/s}^2$. As in Refs.[6,13], $d$ is well approximated by $(d_o^2 H)^{1/3}$, where $d_o$ is the minimum penetration depth for $h = 0$; however, as in Ref.[7], $d$ is also well approximated by $d_o + \alpha |v_o|$. Though the experimental systems of Refs.[6-9] are all different, and were interpreted by four different force laws, they are actually all consistent with the data reported here. This suggests that a single, unified, force law underlies all the observations.

Tsimring and Volfson [14] proposed that the total force on the projectile is

$$\Sigma F = -mg + F(z) + mv^2/d_1, \qquad (1)$$

equal to the sum of gravity plus Coulomb friction plus inertial drag. This is similar to the Poncelet force law, $F_o + cv^2$, used in high-speed ballistics [10,11], but accounts for the depth-dependence of Coulomb friction in granular media. Tsimring and Volfson argued that the form of $F(z)$ should vary from quadratic to constant due to the shapes of the projectile and of the growing crater excavated by its motion. They also showed how such a force-law can approximately account for a depth scaling of $d = (d_o^2 H)^{1/3}$.

The entire dynamics data set of Fig.1 can be used to test the form of Eq.(1). Combining it with Newton's second law, the acceleration at a given fixed depth $z_i$ should be quadratic in speed: $a + g = F(z_i)/m + v^2/d_1$. For each drop height, we therefore examine the acceleration and the speed when the projectile passes through five different fixed depths, $z_i = \{0, -1, -2, -3, -4\} \pm 0.1\,\text{cm}$. The acceleration values are shown vs speed in Fig.3(a), where each point represents a different drop height and where the five colours represent the five fixed depths. The results are quadratic in speed and furthermore, crucially, have the same proportionality factor $d_1 + 8.7 = 0.7$ cm for all five depths. This is demonstrated by the solid curves in the main plot, which become parallel lines when plotted vs speed-squared in the inset. The good agreement shows that the projectile experiences a force $mv^2/d_1$ that is independent of depth. This result can be expressed as $0.8 \rho_g D_b^2 v^2$, and hence can be interpreted as an inertial force required for the projectile to mobilize a volume $D_b^3$ of granular media.

The form of the Coulomb friction-like term may now be examined using the value of $d_1$ deduced above. Since Eq.(1) gives $F(z)/m = a + g - v^2/d_1$, we evaluate the right-hand-side and plot the results vs depth in Fig.3(b). There the curves represent data for different drop heights; the open symbols represent extrapolation to $v = 0$ of $a + g$ data at fixed depth for many drop heights, from Fig.3(a). This produces a good collapse of our *entire* data set, for all drop heights and for all times. It shows that the projectile experiences a force $F(z)$ that is independent of speed and that increases with depth. The form of this force is similar to $F(z)/(mg) = 1 + [3(z/d_o)^2 = 1]\exp(=2|z|/d_1)$, shown in Fig.3(b) as a dotted curve, which is constructed to give $d = (d_o^2 H)^{1/3}$ exactly [15]. This varies from quadratic to linear, and saturates at great depths, just as argued by Tsimring and Volfson. However, the data are better fit over most of the range to a simpler, linear form $F(z) = k|z|$ with $k/m = 1040 \pm 10$ s$^{-2}$, shown by the dashed line. This result can be expressed as $k = 20\mu\rho_g g D_b^2$, which is somewhat larger than expected for ordinary Coulomb friction and observed previously for objects pushed horizontally in a granular medium [16].

Altogether we have shown that a force law $\Sigma F = -mg + k|z| - mv^2/d_1$ can account for the salient features of existing granular impact data. Here the high quality and range of the dynamics data are sufficient to allow the individual depth- and velocity-dependent granular forces to be isolated and demonstrated. The order of magnitude of the latter is $\rho_g D_b^2 v^2$, set by grain density and ball area. The order of magnitude of the former is somewhat larger than $\mu\rho_g g D_b^2 |z|$, set by Coulomb friction. These expressions reveal that the characteristic length scale is given by the diameter of the projectile $L_c = D_b$. This explains why the penetration depth is within an order of magnitude of the ball diameter for a wide range of impact speeds. Since gravity is the only system parameter whose units contain time, the characteristic time and velocity scales must therefore be $T_c = (D_b/g)^{1/2}$ and $V_c = (D_b g)^{1/2}$. Indeed, these explain the typical stopping time and the velocity beyond which the stopping time is constant, seen in the inset of Fig.1b. Also, finally, the force law explains the acceleration discontinuity at stoppage, seen in Fig.1c, as $\Delta a = -g + (k/m)d$. The only puzzles that remain are the linear form of the Coulomb friction term and the precise values of $k$ and $d_1$. In addition to accounting for the full behaviour of the projectile, the fundamental forces demonstrated here to act between projectile and medium must also govern the injection of energy needed to explain spectacular features in the medium such as crater morphologies [4,5] and jet splash heights [17-20].

**Methods**

Spherical glass beads (diameter range $0.25 = 0.35$ mm) are used as a dry non-cohesive granular medium with density $\rho_g = 1.52$ g/cm$^3$, draining angle of repose $\theta_r = 24°$, and friction coefficient $\mu = tan(\theta_r) = 0.45$. A clear plexiglass tube (8-inch outer diameter, 7.5-inch inner diameter, and 12-inch height) is set on a sieve (US sieve size $45$-$60$). The same windbox used in previous experiments [13] is attached under this container. The glass beads are poured into the container to a depth of about 20 cm. The medium is fluidized, and gradually de-fluidized, by a uniform upflow of N$_2$ gas prior to each impact to ensure a homogeneous medium with flat surface. The volume fraction occupied by the beads is $0.590 + 0.004$ after fluidization [21]. A steel sphere of 1-inch diameter is used as a projectile. An acrylic transparent square rod (8 inch long, $1/8 \times 1/8$ inch square cross section) is glued vertically on top of the sphere. A horizontally striped transparent sheet is affixed to one side of the rod. The width and space of the stripes are 0.2 mm. A small metal tip is glued to the top of the rod. The total mass of the projectile plus rod is $m = 69.2$ g. The projectile is held by an electromagnet, which is turned off to commence free-fall of the projectile. A line-scan CCD camera (1024 pixels, 8-bits deep, 50 kHz frame rate) is placed to capture light transmitted through the striped pattern. Images are also acquired before and after each impact for calibration of length scale.

To analyze video data, the velocity is first deduced from the frame rate and the location of the peak in the cross correlation of two successive images of the stripped pattern. Next, the impact time and speed are identified from fits of raw $v(\tau)$ data to $[v_o - g(\tau - \tau_o)]H(\tau - \tau_o) + [v_o + a_o(\tau - \tau_o) + j_o(\tau - \tau_o)^2/2]H(\tau_o - \tau)$, where the fitting parameters are $\{v_o, a_o, j_o, \tau_o\}$ and $H(x)$ is the Heavyside function. The stop time and acceleration discontinuity are identified by fits to $v(t) = \Delta a(t - t_{stop})$ over a small window prior to $t_{stop}$, where $t = \tau - \tau_o$. Position is then deduced from $z(t) = \int_o^t v(t')\,dt'$. The results are consistent with, but more accurate than, the values of $h = v_o^2/(2g)$ and $d = -z(\infty)$ measured directly by a microtelescope mounted to a height gauge. Finally, acceleration is deduced from fits of the velocity data to line segments, where the fitting windows are adjusted so that the acceleration uncertainty is smaller than the larger of 0.5% or $0.005g$. An example of raw video data and analysis is given in the Supplemental Information, along with a discussion of possible effects due to atmospheric air pressure.

This work was supported by the National Science Foundation (DJD) and the Japan Society for the Promotion of Science Postdoctoral Fellowships for Research Abroad (HK).

Correspondence and requests for materials should be addressed to D.J.Durian (djdurian@physics.upenn.edu)

**Figure 1** Complete data set of (a) depth $z(t)$, (b) speed $v(t)$, and (c) net acceleration $a(t) + g$ vs time. The origin is defined by initial impact, and position is measured upwards opposite to gravity. Curves are color-coded according to initial impact speed. Inset to (b): stopping time vs impact speed, along with the characteristic time and velocity scales set by projectile size and gravity. Inset to (c): net acceleration $a + g$ vs depth for the case of zero initial impact speed, along with the expectation $k|z|$ based on Coulomb friction alone.

**Figure 2** Absolute final penetration depth $d$ vs (a) total drop distance $H = h + d$ and vs (b) initial impact speed $v_o = =(2gh)^{1/2}$, where $h$ is the free-fall distance. Symbols are color-coded according to the initial impact speed as in Figure 1. Solid black lines show previously-reported empirical scaling rules: (a) $d_0 = (d_o^2 H)^{1/3}$, where $d_0$ is the fitting parameter [6,13]; and (b) $d \pm d_o = \alpha|v_0|$, where both $d_0$ and $\alpha$ are fitting parameters [7].

**Figure 3** (a) Net acceleration $a + g$ vs velocity, at five specific fixed depths $z_i$. Curves represent fits to $a + g = F(z_i)/m + v^2/d_1$, where $d_1$ is set to 8.7 cm and $F(z_i)/m$ is adjusted for each depth. The inset shows the same plot but vs $v^2$, in order to demonstrate the quadratic dependence on speed and the constancy of $d_1$. (b) Net acceleration $a + g = v^2/d_1$ vs depth $z$; according to Eq.(1), this should collapse all data to the friction force $F(z)/m$. The curves are color-coded according to the initial impact speed as in Figure 1; open symbols represent the fitting results $F(z_i)/m$ from (a). The dashed and dotted curves represent candidate forms for $F(z)/m$ vs $z$.

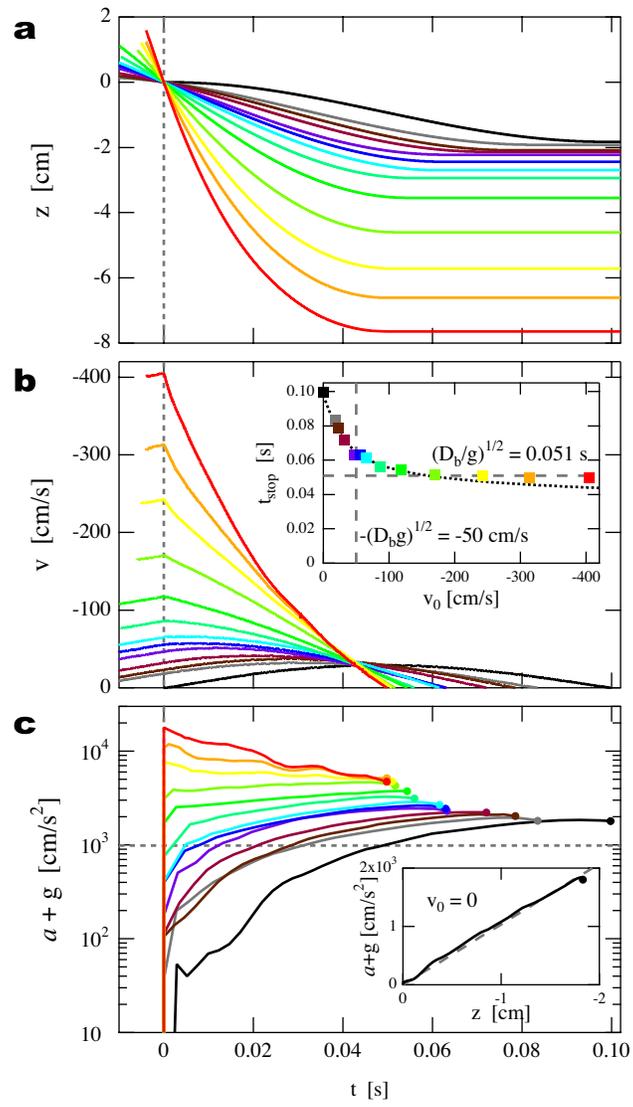

Figure 1

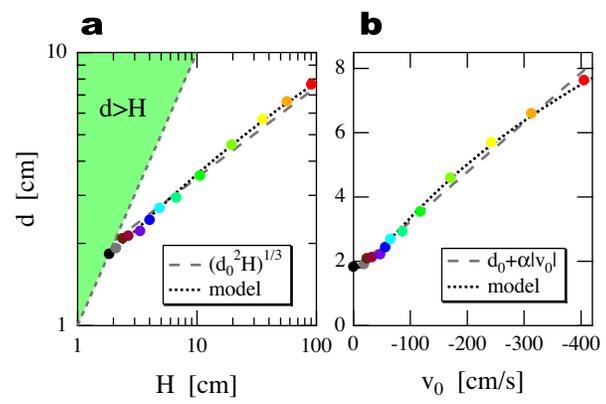

Figure 2

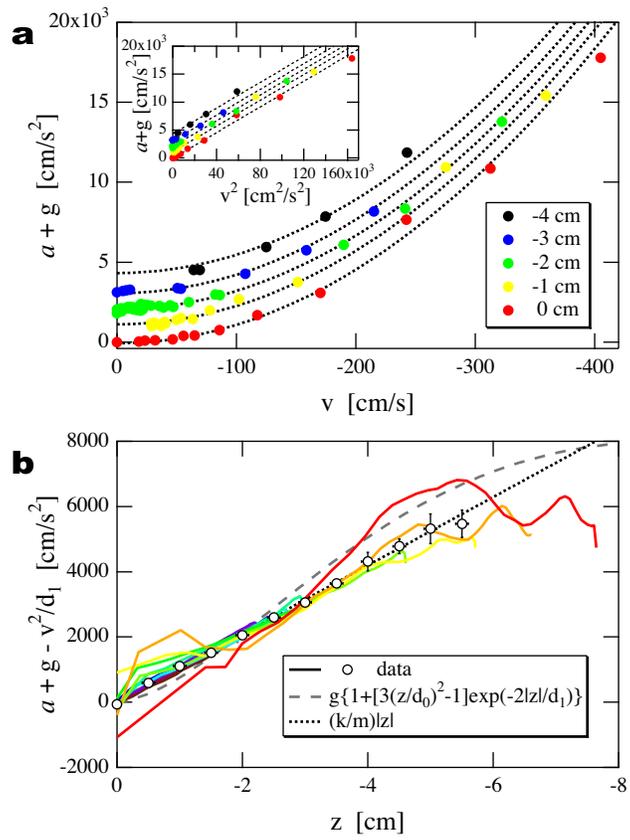

Figure 3

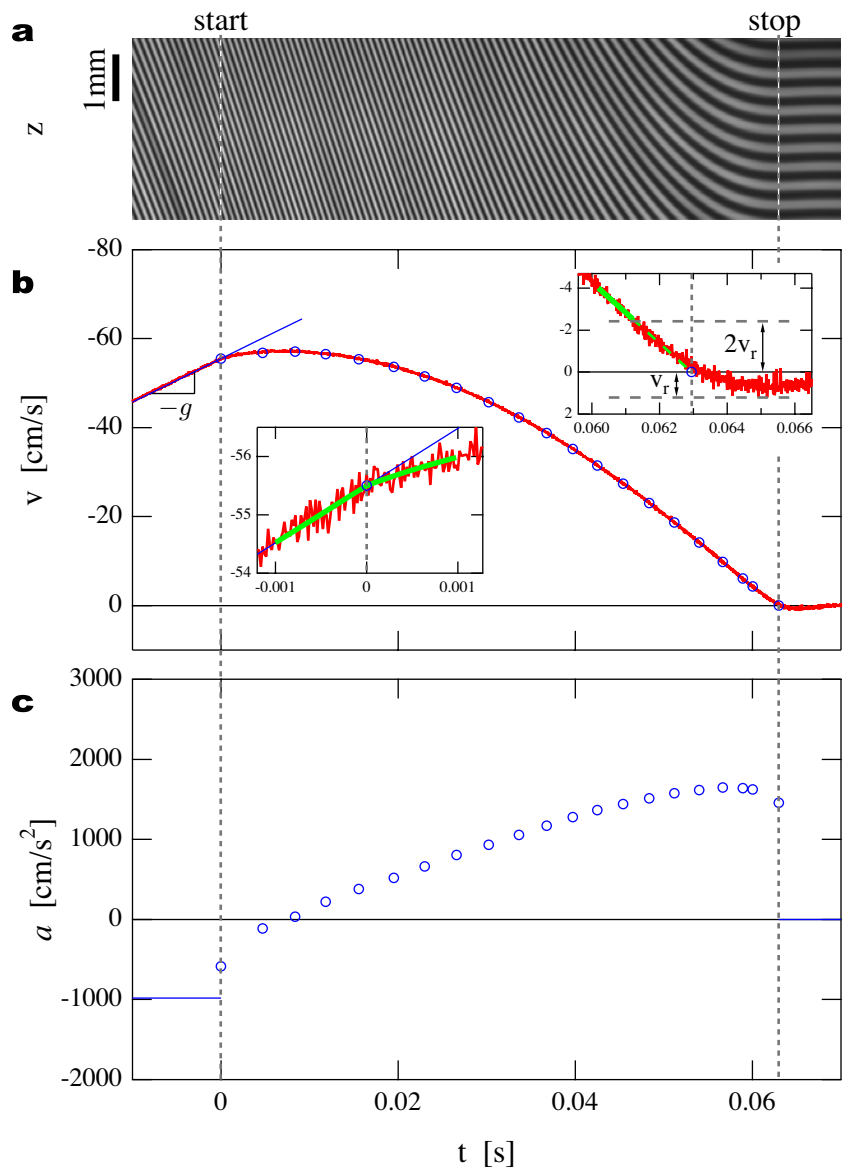

Supplemental Figure 1. Example of data and analysis as described in Methods: (a) images from line-scan camera, (b) projectile velocity, and (c) projectile acceleration, all vs time. In (b) the red curve represents velocity data computed from the video images, the blue line has slope $-g$ indicating free-fall, and the open blue circles represent the average velocity from the straight-line segments used to deduce the acceleration results shown in (c). The left inset in (b) is a blow-up of velocity vs time at initial impact, identified by the fit shown in green. The right inset in (b) is a blow-up of velocity vs time as the ball comes to rest. Note that the velocity data exhibit damped oscillations, due to slight flexing of the brass mesh sieve that supports the granular medium (see Supplemental Figure 2). Since it is not obvious when the projectile comes to rest with respect to the moving medium, we identify the stop time by linear extrapolation, shown in green, of velocity data greater than twice the maximum rebound speed. Note that the velocity resolution of 0.5 cm/s is evident as the scale of noise in the insets of (b).

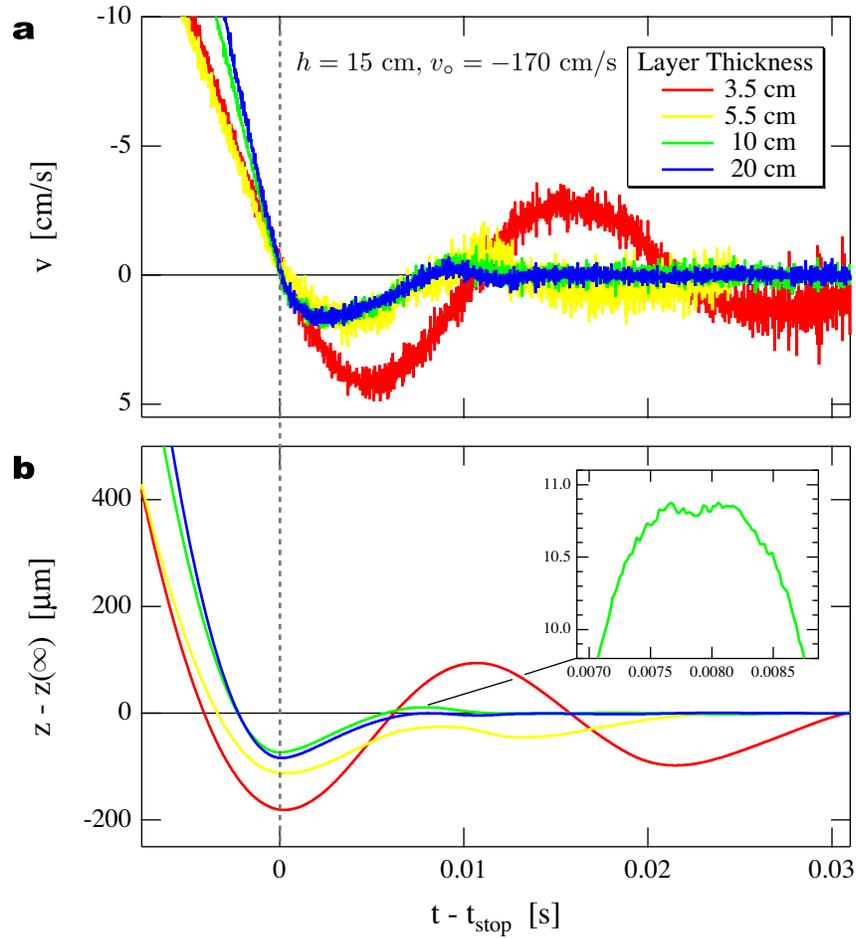

Supplemental Figure 2. (a) Velocity and (b) position vs time for four runs with the same drop height, but with layers of granular medium of different thickness. The observed oscillations decrease with increasing layer thickness, L, but are still present in the thickest layers studied, where the extent of motion is of order 100 μm. This trend and scale are both counter to expectation based on compression of the medium, $\Delta L = LP/Y$, which gives 3 nm for pressure $P = mg/D_b^2$ and Young's modulus for glass $Y = 70$ GPa. Rather, the observations are consistent with flexing of the brass mesh sieve, which is expected to decrease with layer thickness and to have scale 300 μm set by deflection measured under the weight of the projectile alone. Note that the position resolution of 0.1 μm is evident as the scale of noise in the inset of (b).

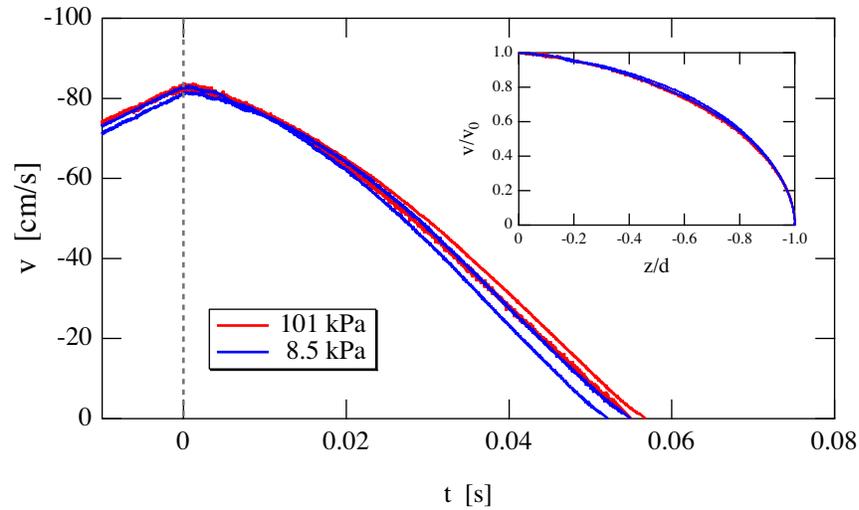

Supplemental Figure 3. Velocity vs time (main) and normalized velocity vs position (inset) for runs with the same drop height, but with and without the presence of air. To within reproducibility, the results in air (101 kPa) are no different from the results in vacuum (8.5 kPa). By contrast, such a change in pressure dramatically affects the granular jet for deep impact [20].